\documentstyle[11pt]{article}
\makeatletter
\let\chapter\hid@chapter
\makeatother
\def\lsim{\lower.5ex\hbox{$\; \buildrel < \over \sim \;$}}
\def\gsim{\lower.5ex\hbox{$\; \buildrel > \over \sim \;$}}
\begin{document}
\pagenumbering{arabic}
\title{VLF observation during Leonid Meteor Shower-2002 from Kolkata}

\author{Sandip K. Chakrabarti$^{1,2}$, S. Pal$^{2}$, K. Acharya$^{2}$, S. Mandal$^{2}$, 
S. Chakrabarti$^{2,3}$,\\
R. Khan$^{2,4}$, B. Bose$^{2}$}

\maketitle

\noindent 1.  S.N. Bose National Centre for Basic Sciences, JD-Block, Salt Lake, Kolkata 700098\\
\noindent 2.  Centre for Space Physics, P-61 Southend Gardens, Kolkata, 700084\\ 
\noindent 3. Maharaja Manindra Chandra College, 1 Ramkanta Bose St., Kolkata, 700002\\
\noindent 4. Bidhan Nagar High School, BD block, Salt Lake, Kolkata, 700064\\

\begin{abstract}
Using a Gyrator-II Loop antenna tuned at 19.0Khz, we monitored the meteor shower
during 17-24th November, 2002. We observe the primary peak
at 3h58m (UT) on the 19th of November, 2002. We distinctly observed several
`beadlike' and `exponentially dropping' signals. The `beadlike' signals were
more in abundance on the 18th of November, 2002, one day prior to the actual encounter.
\end{abstract}

\noindent PUBLISHED IN INDIAN JOURNAL OF PHYSICS, 2002, 76B, 693

\section{Introduction}

Very Low Frequency (VLF) project of CSP has been monitoring VLF activities for quite some 
time and has detected solar flares as well. During 
the recent Leonid shower event the CSP antenna was tuned to 19KHz and 
continuous observations were made for seven days. In this Rapid Communication
we present the first report of this observation.

Leonid meteor showers are observed around 16-19th November for 3-4  successive years 
after the perihelion passage of the periodic comet Temple-Tuttle ($T\sim 33$ Yrs). 
On February 28, 1998, the comet reached its perihelion and in 1998-2002 the
visual observations indicated very good meteor showers. When the
earth's orbit crosses  the debris left over by the
comet on its path in its previous passages, a shower or storm may form, depending 
on the intensity of events. In 2002, several calculations indicated that 
there would be two peaks, one  would be from Europe at around 4.00UT (due to earth's
passage through the dust trail left behind in 1767 A.D.) and the other
would be seen from North America at around 10.40UT (due to earth's
passage through the dust trail left behind in 1866 A.D.) [2]. Apart from encountering
a rare celestial event, scientists are interested in Leonids because of the 
prospect of detecting bio-molecules which are thought to have contaminated the
earth after being produced in space [3-5].
Furthermore, it has been reported that electrophonic sounds have been
recorded during Leonid showers [6-7]
which are thought to be due to the fact that Very Low Frequency (VLF)
electromagnatic (EM) waves are produced during the passage of a fire-ball. 
Keay [8] and later Bronshten [9] suggested that these VLFs could be
produced due to entangling of earth's magnetic field in the tail of the bow-shock
generated by an incoming meteor. More recently, it is thought that the VLFs are
produced when a fireball bursts. In any case, the EM radiation emitted 
is found to have frequency ranging from a few Hz to about $30$KHz [8]. 
This range is divided into three parts (a) Ultra-Low-Frequency (ULF, 
$\nu <300$Hz), (b) Extreme Low Frequency (ELF,
$0.3 <\nu< 3$ KHz) and (c) Very Low Frequency (VLF, $3 < \nu< 30$ KHz).

In Centre for Space Physics, a team of scientists are engaged in monitoring 
solar activities continuously using two  VLF detectors, one located near Kolkata
and the other is located at Malda. Reports on these would be made elsewhere.
Presently, we consider only the results of the monitoring of 2002 Leonid meteor shower.

\section{Experimental Setup}

The loop Antenna is made of a square frame of one meter on each side and several turns of 
shielded single core wire is used to receive the signal. The signal is then
amplified and is fed into the audio card of a Pentium-IV computer located inside the
laboratory. The audio signal is 
sampled at $3.2$ times per second. The antenna is tuned at $19$KHz, away from the nearest 
$18.2$KHz signal transmitted by VTX3, Indian Navy traffic station at Vijayananarayanam. 
It is aligned along the East-West direction. The magnetic field of the VLF signal 
induces a current in the antenna. The antenna near Kolkata was placed at a height
of about $12$ meters from the ground. The gain of the receiver is adjusted to obtain  
a decibel level of around $1500$ when there is no signal.

\section{Results}

Figs. 1(a-c) show  results of the output from 2.15am to 12.15am (-3h15m to 6h45m UT)
on (a) 18th, (b) 19th and (c) 20th of November, 2002. The dates on each curve
are marked. The otherwise steady result is perturbed due to passages of meteors,
and possible atmospheric phenomena such as thunderbolts [6, 11-12]. However,
the days had very clear skies and no serious `thunder-bolt' related events 
were expected. On the 19th of November,  at 3h58m UT (9h28m IST), 
there was a distinct peak comprising of about $70$ sub-structures,
presumably from individual strong events.  (Expected visual rate [Zenithal hourly rate
or ZHR] Kolkata area was around $100$ per hour at this time,
while the visual observation of the CSP team from Bolpur was $25$ per hour in between
-1h30m to -0h30m UT). On the 20th November there was some enhanced activity
at around 4h27m UT (9h57m IST) but not as much as on the 19th.
It is to be noted that the visual peak was observed by Leonid-MAC team at 4.09UT [3]
very close to the peak found in VLF although several reports [10]
indicated that the peak lies between 3h48m to 4h04m UT thereby bracketing
our observation.

In Fig. 2, we present the results during the peak from 3h30m UT to 
4h30m UT. The duration of the shower seems to be from 3h38m to
4h28m, i.e. the shower lasted for about $50$ minutes.
This is similar to what is reported by Leonid-MAC and IMO. Each sub-structure
in the peak is similar to an exponential decay curve,
but the enhancement of the base (see Fig. 1b and Fig. 2) indicated that 
new signals were injected even before the earlier one has time to decay completely.
The signal is easily modeled by $\Sigma_i a_0(i) exp(t/t_0)$ over the injected 
We also observed two secondary peaks at around 5h UT and 5h45m UT.
Some disturbances have been observed but no distinct peak was found
at around 10h50 UT when the American peak was supposed to be formed. 
It is possible that the VLF signal at $19$KHz is washed out while traveling 
half-way across the globe.

Throughout our observation there were mainly two types of signals. One is 
`bead'-like and the other is exponentially decaying. In Fig. 3a, we show
one example of the `bead-like' signal, typical of the profile of multiple
meteors. However unlike earlier observations [6-7, 12]
where the signals lasted a few tenths of a second, 
our observed  duration was much longer, several minutes. Also,
earlier observers showed sharp spike like  features at the 
beginning, while we miss it during the `bead-formation'. It is 
possible that due to our low time resolution ($\sim 300$ms), multiple
events merged together and long lasting beads were formed. In Fig. 3b,
we demonstrated a typical signal dropping exponentially. 

\section{Concluding Remarks}

The VLF project of CSP has been able to observe the peak very distinctly 
at about 3h58m UT on the 19th of November, 2002. We made the observation
at a frequency far away from previously reported observations during 
1998 and 1999 showers and confirm that the meteors do emit VLF
signals even at 19KHz during their entry in earth's atmosphere.

What could be the cause of the VLF emission and what is the range in which
it is emitted? We believe that the bow-shock that is formed in-front of the 
highly supersonic meteor becomes unstable due to Kelvin-Helmholtz (K-H) instability
along with the tangential discontinuity which separates the evaporated 
matter from the meteor head and the shock-compressed matter in between the
bow-shock and the tangential discontinuity. If one considers the 
bow-shock alone, a strong shock will compress the flow 
by a factor of $\rho_1/\rho_2\sim 4$ [13] and the tangential velocity difference would be
$v_1-v_2 \sim 0-30$km depending on the location of the bow-shock, highest being
at an angle $\theta \sim 30-45^o$ with the propagation axis
and the lowest being at the stagnation point ($\theta \sim 0$) and downstream farther away
($\theta \sim 180^o$), where the bow-shock loses its identity. The frequency 
$\nu$ of the K-H instability is given by [13],
$$
\nu^2_{KH} = \frac{1}{4\pi^2} \frac{\rho_1\rho_2}{(\rho_1+\rho_2)^2} (v_1-v_2)^2 .
$$
Assuming, $\rho_1/\rho_2\sim 1/4$, we find that anywhere between $0$ to $180$ KHz could
be produced with very small amplitude on both the ends (at the stagnation point
of the bow-shock and farther out). It is possible that the earth's magnetic field
entangled in the vortices at this K-H unstable interface generate E-M waves of 
the same frequency.

\newpage

\centerline{Figure Captions}

\noindent Fig. 1(a-c): VLF signal variation during 2h15m  and 12h15m IST (-3h15m UT to 6h45m UT)
during 18th-20th November, 2002. The peak occurs at 3h58m UT on the 19th.
Some `beadlike' and `exponentially decaying'
signals could also be found.

\noindent Fig. 2: Details of the signal during the peak hour of the Leonid shower 
on the 19th of Nov. 2002. The peak is made up of at least $70$ distinct events and is observed
at 3h58m UT hour with a duration of about 50 minutes. There were secondary peaks at 5h UT
and 5h45m UT.

\noindent Fig. 3(a-b): (a) Details of a `beadlike' signal lasting for about three minutes. It is 
possible that it is made up of superposition of smaller events. (b) Details of an
`exponentially decaying' event. The recovery time scale is about $100s$.

\end{document}